\documentclass[namedreferences]{SolarPhysics}
\usepackage[optionalrh]{spr-sola-addons} 
\usepackage{graphicx}                    
\usepackage{color}                       
\usepackage{url}                         

\newcommand{\etal}{{\it et al.}}

\begin{document}

\begin{article}

\begin{opening}

\title{Low-Energy Cutoffs In Electron Spectra Of Solar Flares: Statistical Survey}


\author{E. P. \surname{Kontar}$^{1}$\sep
 	E. \surname{Dickson}$^{1}$\sep
	J. \surname{Ka\v{s}parov\'{a}}$^{2}$
}

%
\runningauthor{Kontar, Dickson and Ka\v{s}parov\'{a}}
\runningtitle{Low-Energy Cutoffs in Electron Spectra}

%
  \institute{$^{1}$ Department of Physics \& Astronomy, University of Glasgow, G12 8QQ, UK\\
	   $^{2}$ Astronomick\'{y} \'{u}stav AV \v{CR}, v.v.i., Fri\v{c}ova 298, 251 65 Ond\v{r}ejov, Czech Republic
}

\begin{abstract}
The Ramaty High Energy Solar Spectroscopic Imager (RHESSI) X-ray data
base (February 2002 -- May 2006) has been searched to find solar
flares with weak thermal components and flat photon spectra. Using
a regularised inversion technique, we determine the mean electron
flux distribution from count spectra of a selection of events with
flat photon spectra in the 15--20~keV energy range. Such spectral
behaviour is expected for photon spectra either affected by photospheric
albedo or produced by electron spectra with an absence of electrons
in a given energy range, e.g. a low-energy cutoff in the mean electron spectra
of non-themal particles. We have found 18 cases which exhibit a statistically
significant local minimum (a dip) in the range of 10--20 keV.
The positions and spectral indices of events with low-energy cutoff
indicate that such features are likely to be the result
of photospheric albedo. It is shown that if the isotropic albedo
correction was applied, all low-energy cutoffs in the mean electron spectrum
were removed and hence the low energy cutoffs in the mean electron
spectrum of solar flares above $\sim$12 keV cannot be viewed as real features in the electron
spectrum. If low-energy cutoffs exist in the mean electron spectra,
the energy of low energy cutoffs should be less than $\sim$12 keV.
\end{abstract}
%

\end{opening}

%

\section{Introduction}

X-ray observations are often used to infer various properties
of energetic electrons accelerated during the solar flares.
The spatially integrated X-ray photon spectrum
$I(\epsilon)$ (photons~cm$^{-2}$~s$^{-1}$~keV$^{-1}$) is related
to mean electron flux spectrum ${\overline F}(E)$ (electrons
cm$^{-2}$~s$^{-1}$~keV$^{-1}$) via the rather simple linear
integral relation \cite{Brown03}
\begin{equation}
\label{Idef} I(\epsilon) = \frac{{\bar n} V } {4\pi R^2}\int_\epsilon^\infty
{\overline F}(E) Q(\epsilon,E)dE
\end{equation}
for source volume $V$, mean plasma density $\bar n$, and
isotropic bremsstrahlung cross-section per unit photon energy $\epsilon$,
$Q(\epsilon,E)$ \cite{Haug97}. Although the angular distribution
of energetic electrons is generally unknown, the recent observations \cite{kontarbrown06}
suggest rather close to isotropic distribution of electrons.
The exact plasma density distribution and flaring volume are also poorly known
and therefore value ${\bar n}V{\bar F}(E)$ is normally inferred.
The value ${\bar n}V \bar F(E)$ is model independent
\cite{Brown03}, and the detailed energy
structure of this is related to electron acceleration
and propagation physics. Radio emission spectrum of solar energetic particles,
although normally only at above a few hundred keV energies, is an alternative approach
to infer electron beam and plasma parameters \cite{Altyntsev08}.

The spatially integrated spectrum of energetic electrons ${\bar n}V \bar F(E)$
is often {\it approximated} as a sum of a isothermal Maxwellian distribution and
a non-thermal power-law distribution \cite{Holman03}.
The thermal component often dominates the overall spectrum
at low energies $\leq 20$~keV and little can be said about
the low-energy part of non-thermal distribution.
However, the low-energy part of non-thermal spectra plays a
crucial role in the solar flare diagnostics. Most of the
non-thermal electron energy is concentrated in this part, hence this
defines the total energy budget of the flare. In addition,
this part of the spectrum is more effectively influenced by
various electron propagation effects like collisions \cite{Brown71} or
beam-plasma interactions \cite{melnik99}, thus playing
an important role in the electron transport diagnostics
in the solar flares. Various model-based methods to find the value of low energy
cutoff have been used. Requiring that
the assumed thermal emission dominate over non-thermal
emissions \inlinecite{sui05} find a low energy cutoff of $\sim 24$ keV should be present.
Assuming "theoretical Neupert effect" to be satisfied \inlinecite{veronig05}
conclude that the low energy cutoff should be between $10$ keV and $30$
keV for four flares analysed in the paper.
\inlinecite{Hannah08} have used empirical
relationship between the observed parameters of the photon
power-law fit and the low-energy cutoff of the electron distribution
and have found that the low-energy cutoffs in microflare events
could range from $9$ to $16$~keV with the median being around $12$~keV.
In this paper we will focus on the {\it model-independent} inference of
low-energy cutoffs in the mean electron spectra.

High resolution spectra observed by RHESSI \cite{lin02}
allows us to infer detailed structure of electron distribution
often never seen before. \inlinecite{piana03} have demonstrated that
the mean electron spectrum ${\bar n}V{\bar F}(E)$ has a statistically
significant local minimum at approximately $50$ keV
in the electron spectrum of GOES X-class July 23, 2002 solar flare,
although this feature is likely to be an instrumental effect caused
by a pulse pile-up. \inlinecite{kontarbrown06} show
that some electron spectra inferred from RHESSI X-ray
spectra free from pile-up issues seem inconsistent
with a simple collisional thick-target model \cite{Brown71}.
However, the photon spectra of these events should be corrected
for albedo - Compton back-scattered X-rays \cite{kontar06}.
\inlinecite{kasparova05} have shown that the spectrum of the
August 20, 2002 event with a puzzlingly large value of the low-energy
cutoff $\sim 30$ keV can be understood in terms of the photospheric albedo.

Flares showing a weak thermal component allow us to scrutinize
the low-energy part of the non-thermal distribution of
electrons. The analysis can be done either by {\it assuming}
a functional form of the electron spectrum \cite{Holman03}
or by using the regularised inversion techniques \cite{kontar04}.
It is known that flat X-ray spectra (low value of photon spectral index) can require
low-energy cutoffs in the power-law distributions when a functional
form is assumed \cite{kasparova05,sui07},
whereas the model independent approach, via the regularised inversion technique,
\cite{piana03} may show a dip or a gap in the electron
distribution \cite{kontar06,kasparova07}.

In this paper we present the results of a systematic search for dips
in the mean electron flux distribution using the RHESSI solar flare
database  for the period of Feb 2002 - May 2006. Section~\ref{analysis} describes
the selection criteria for the flare photon spectra and the application
of the regularised inversion method for the determination of the corresponding
mean electron flux spectra. Section~\ref{dips} discusses energies and depths of the obtained
statistically significant dips and their relation to the photospheric albedo.
The analysis confirms previous suggestions that the isotropic albedo correction
is capable of removing all statistically significant dips in the mean electron
flux distribution. The obtained results are summarised in Section~\ref{summary}.

\section{Data analysis}\label{analysis}

As a basis, we used the list of 398 flares with weak thermal component
previously determined by \inlinecite{kasparova07}. Although this has limited
the total number events for our analysis, it has helped us to
avoid various effects, such as pulse pile-up and particle contamination,
complicating the spectral analysis \cite{Schwartz02}.
Next, we chose the 177 events with the smallest values of spectral
index $\gamma_0 \leq 4.0$, where $\gamma_0$ was measured in the range
between 15 and 20 keV - see \inlinecite{kasparova07}.

For each flare, the spectra were accumulated over the duration
of the impulsive phase, i.e. in the interval when counts at
energies above 50 keV were above background (Figure 1).
The spectra were generated in the energy range from 3 to 100 keV
with 1 keV resolution avoiding detectors 2 and 7 due to their
low resolution \cite{smith02}. The background counts were
removed in a standard way \cite{Schwartz02}.

To obtain a starting point for the regularised inversion,
spectra were forward fitted assuming an isothermal plus a non-thermal
double power-law distribution of ${\bar F(E)}$, for example \inlinecite{Holman03}.
Spectra were then inverted within
OSPEX\footnote{\url{http://hesperia.gsfc.nasa.gov/ssw/packages/spex/doc/ospex_explanation.htm}}
using the regularised inversion routines\footnote{\url{http://www.astro.gla.ac.uk/users/eduard/rhessi/inversion/}} \cite{kontar04} minimizing the functional \cite{tikhonov07}

\begin{equation}\label{mproblem}
\mathcal{L}(\bf{\overline{F}})\equiv \|{\bf{A}}{\bf{\overline{F}}}-{\bf{C}}\|^2+ \lambda
\|{\bf{L\overline{F}}}\|^2=\mbox{min}
\end{equation}
where
\begin{equation}\label{eq:A}
{\bf A} = {\bf R}\,{\bf B}
\end{equation}
\begin{equation}\label{eq:B}
B_{ij} = \frac{{\overline{n}}V}{4\pi R^2}
Q((\epsilon_{i+1}+\epsilon _{i})/2,(E_{j+1}+E_{j})/2) \, \Delta
E_j
\end{equation}
where ${\bf R}$ is a spectral response matrix of RHESSI converting
photons to photon counts, ${\bf B}$ is a matrix representation of our linear
integral (\ref{Idef}), ${\bf L}$ is the matrix representation of the additional
constraint,  ${\bf C}$ is data vector of background-subtracted
count spectrum  (counts~cm$^{-2}$~s$^{-1}$~keV$^{-1}$), and ${\overline{\bf{F}}}$ is the vector
of unknown density-weighted mean electron spectrum ${\bar n}V{\overline{F}}$.

The equation \ref{mproblem} can be solved analytically using
Generalised Singular Value Decomposition. Regularisation parameter,
$\lambda$, is determined from the analysis
of normalized residuals, $r_k=(({\bf{A}}{\overline{\bf{F}}})_k - {\bf{C}}_k)/\delta {\bf C}_k$,
where $\delta {\bf C}_k$ are the uncertainties of the count spectrum.
Then the deviation weighted by the error
\begin{equation}\label{chi2}
\|({\bf{A}}{\overline{\bf{F}}}_{\lambda} - {\bf{C}})(\delta {\bf C})^{-1}\|^2 = \alpha
\end{equation}
accounts quite accurately for point-to-point error variation.
Indeed $\lambda $ defined by Equation (\ref{chi2}) has accounted
for detailed structure of errors. Parameter $ \alpha$ is chosen
to make the residuals $r_k$ to be close to gaussian \cite{kontar04}.

Using first order regularisation, i.e. operator ${\bf L}$
being a finite difference representation of a derivative operator,
we found mean electron flux spectrum ${\bar n}V{\bar F}(E)$ for all 177
flares with spectral index $\gamma _0$ less than 4.

\section{Local minima (dips) in the mean electron flux spectrum}\label{dips}

With the mean electron flux determined, the spectrum was examined
for local minima or so-called dips. These dips were analyzed
to infer the dip (local minimum) parameters: the energy $E_{\rm d}$ at which
the dip minima occurs and the depth of the dip $d$ in terms of $\sigma$,
where $\sigma$ is the statistical uncertainty on the inferred
mean electron spectrum ${\bar n}V{\bar F}(E)$. This depth was calculated by dividing
the difference between the minimum and the maximum above the dip
in units of electron spectra uncertainty at the minimum (Figure 2).
We have found 18 events with a dip depth deeper than 1 $\sigma$ in the electron
distribution function. The details of these events are presented in Table 1.
Some of the events presented in the Table 1 have been found
using thick-target model fit with a single power-law and
low energy cutoff \cite{sui07}.

The local minima in the mean electron spectra are 6-10 keV wide and hence cover
a few statistically independent energy points. For example, if a dip is three
points wide at $1\sigma$ level in each point, the probability to find three
consecutive points outside $1\sigma$ interval is $(1-0.68)^3 = 0.03$
and the corresponding statistical significance of the minimum is $1-0.03=0.97$.
In general, given that the errors have normal distribution the statistical
significance of the local minimum is $1-\Pi _{i=1}^{N}(1-\mbox{erf}(d _i/\sqrt{2}))$,
where $N$ is the total number of statistically independent
points in a dip (local minimum) and $d _i$ is the depth of each point
in units of the corresponding $\sigma _i$ uncertainties. The sizes of statistically
independent energy bins can be estimated from horizontal errors (Figure 2).
Thus, the local minimum (dip) shown in Figure 2 has statistical
significance $\sim 1-[1-\mbox{erf}(2.9/\sqrt(2))][1-\mbox{erf}(1.2/\sqrt(2))]\approx 99.9\%$.

\begin{table}
\caption{Events with a local minimum (dip) in the mean electron spectrum
dipper than 1$\sigma$; $d_i$ is the depth of a dip in $\sigma$; $E_d$ is the energy of
the local minimum; $\mu= \cos(\theta)$ is the cosine of flare heliocentric angle; $\gamma _0$
is the photon spectral index measured in the range $15$ - $20$ keV, with typical uncertainty
$\pm 0.2$.}
\begin{tabular}{ | c | c | c | c | c | c |  }
\hline
Flare Date &  Time & $d_i$ ($\sigma$) & $E_d$ (keV) & $\mu$ & $\gamma_0$\\
\hline
 11-Apr-2002 &	03:06:08.00 &	1.7 &	15.5 &	0.96 &  1.6	\\
 25-Apr-2002 &	05:55:12.00 &	2.5 &	16.5 &	0.96 &	1.7	\\
 29-Jun-2002 &	09:29:40.00 &	2.0 &	15.5 &	0.16 &	2.7	\\
 30-Jul-2002 &	17:37:36.00 &	1.9 &	18.5 &	0.97 &	2.1	\\
 17-Sep-2002 &	05:51:12.00 &	2.7 &	16.5 &	0.74 &	1.7	\\
 24-Oct-2002 &	00:09:24.00 &	2.0 &	15.5 &	0.94 & 	2.2	\\
 22-Nov-2002 &	13:29:36.00 &	2.8 &	17.5 &	0.93 & 	2.5	\\
 10-Mar-2003 &	10:02:56.00 &	1.2 &	13.5 & 	0.72 &	2.9	\\
 20-Nov-2003 &	05:10:36.00 &	1.3 &	12.5 & 	0.93 &	2.9	\\
 1-Apr-2004 &	23:00:32.00 &	2.9 &	15.5 &	0.88 &	2.6	\\
 20-May-2004 &	17:16:12.00 &	1.4 &	15.5 &	0.19 &	2.9	\\
 19-Jul-2004 &	20:56:52.00 &	1.9 & 	16.5 &	0.73 &	2.0	\\
 14-Aug-2004 &	08:15:30.00 &	1.9 &	18.5 &	0.82 &	1.6	\\
 28-Oct-2004 &	12:13:32.00 &	1.5 & 	16.5 &	0.30 &	3.1	\\
 9-Nov-2004 &	15:10:08.00 &	1.1 &	15.5 &	0.66 &	3.6	\\
 30-Nov-2004 &	03:56:12.00 &	1.2 &	14.5 &	0.97 &	2.7	\\
 21-Jan-2005 &	06:32:20.00 &	1.0 &	15.5 &	0.29 &	2.5	\\
 5-Apr-2006 &	22:45:28.00 &	2.5 &	17.5 &	0.62 &	2.2	\\
\hline
\end{tabular}
\end{table}

\begin{figure}
\begin{center}
\includegraphics[width=85mm]{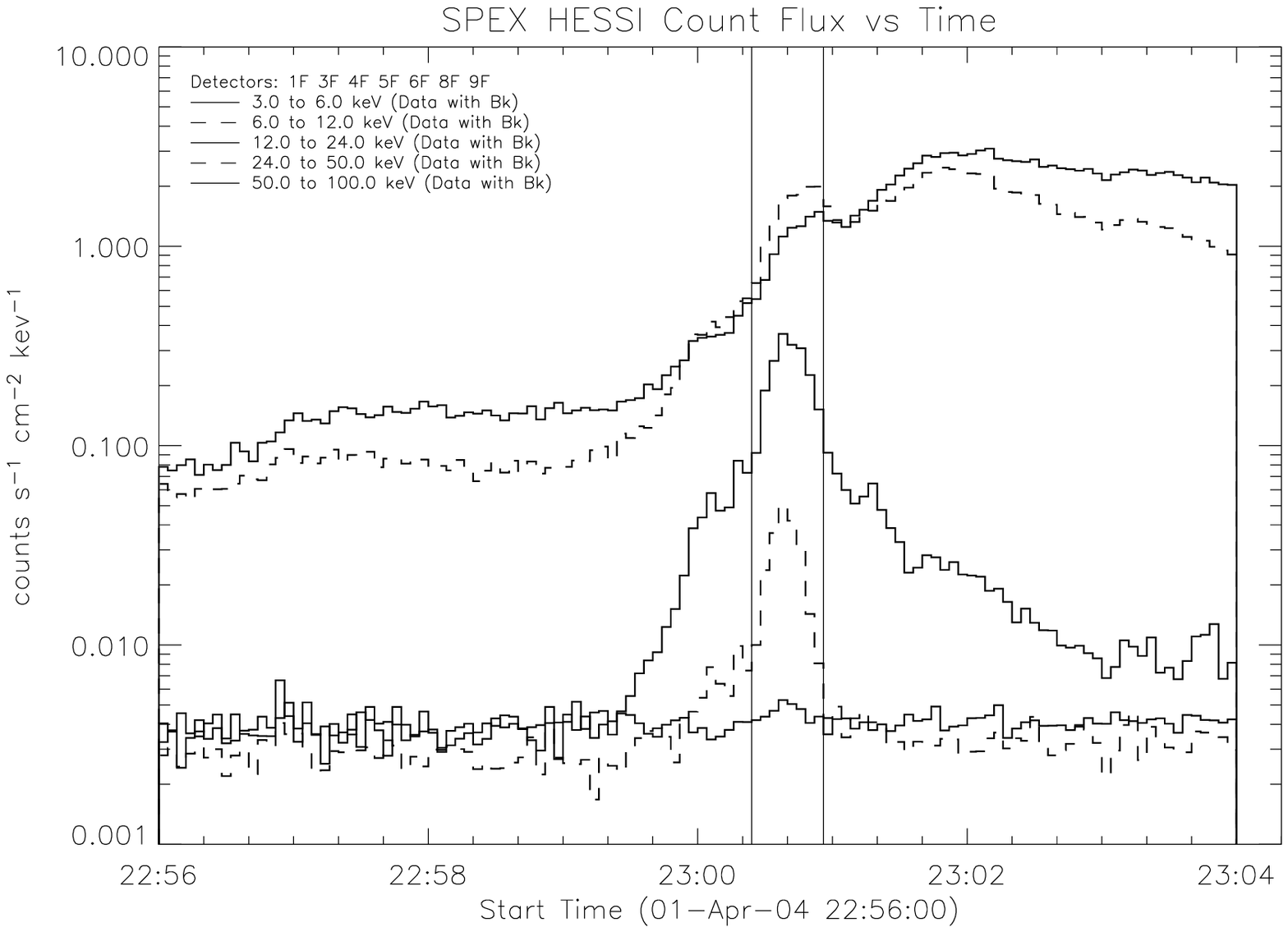}
\includegraphics[width=87mm]{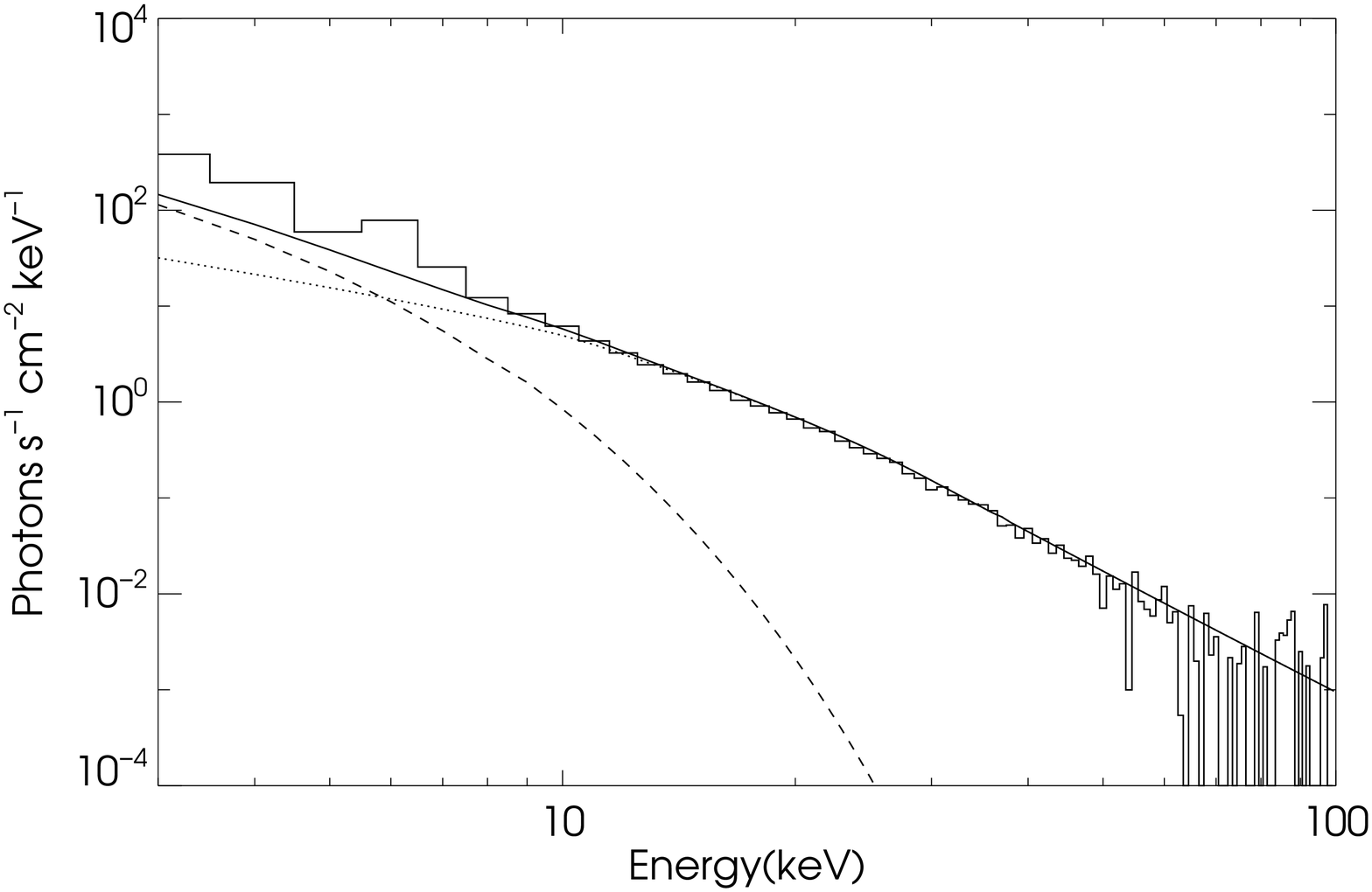}
\end{center}
\caption{Example of a solar flare with flat
photon spectrum. Upper panel: RHESSI light curves; The vertical lines show the accumulation time
interval for spectroscopic analysis. Lower panel: Photon spectrum and forward fit
(solid line), isothermal component (dashed line), nonthermal component (dotted line).}
\label{fig1}
\end{figure}

\begin{figure}
\begin{center}
\includegraphics[width=80mm]{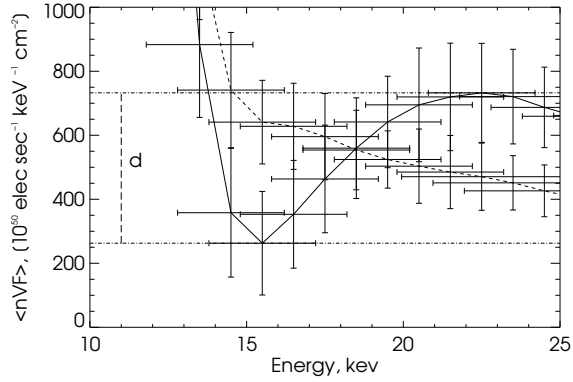}
\end{center}
\caption{Mean electron distribution spectrum
for April 1, 2004 $\sim 23:00$~UT solar flare. The observed
electron spectrum (solid line) and electron spectrum after isotropic albedo
correction (dashed line) are given with $1\sigma$ error bars.
The dip depth, $d$, is shown. }
\label{fig2}
\end{figure}

\begin{figure}
\begin{center}
\includegraphics[width=49mm]{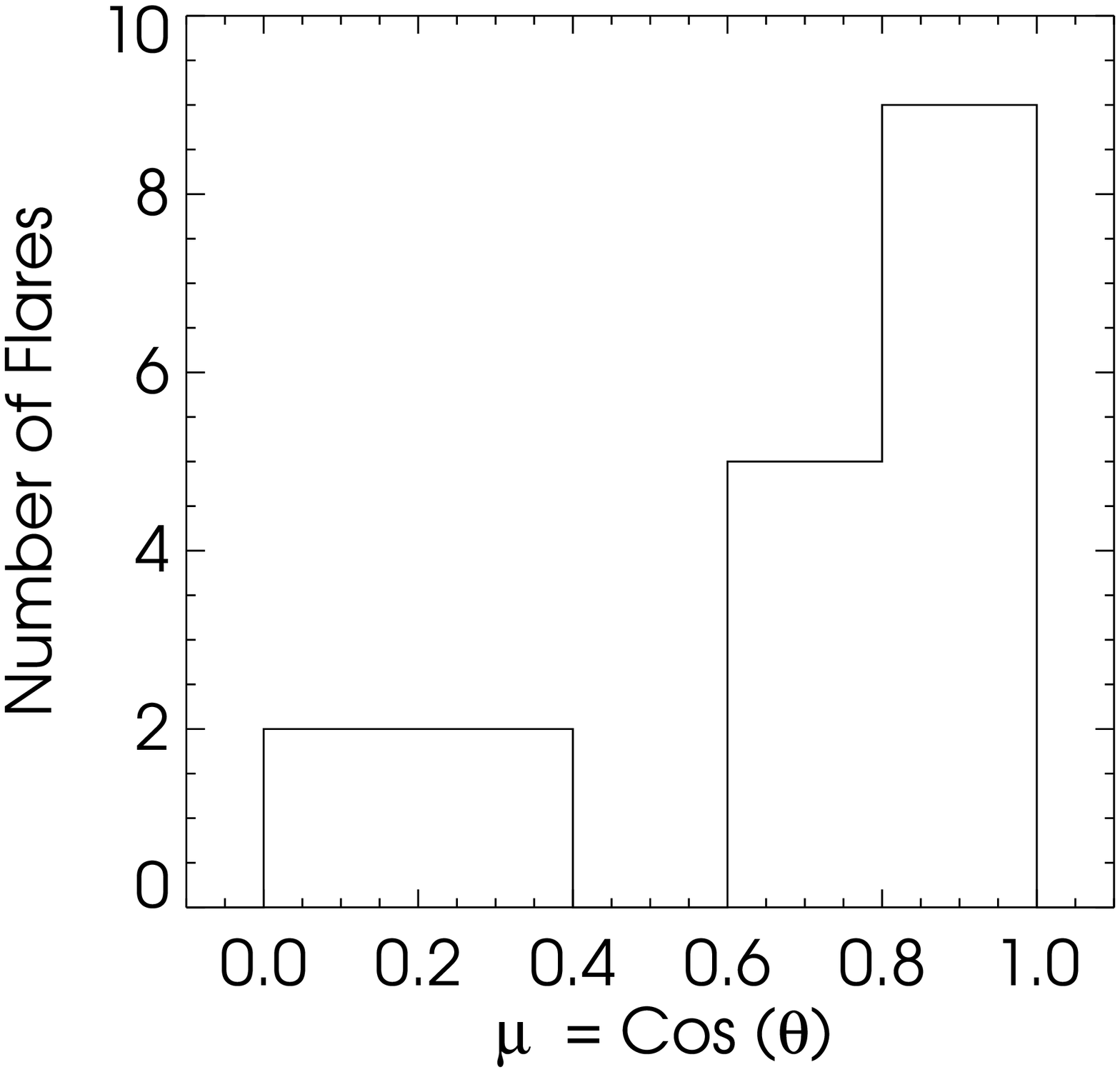}
\includegraphics[width=49mm]{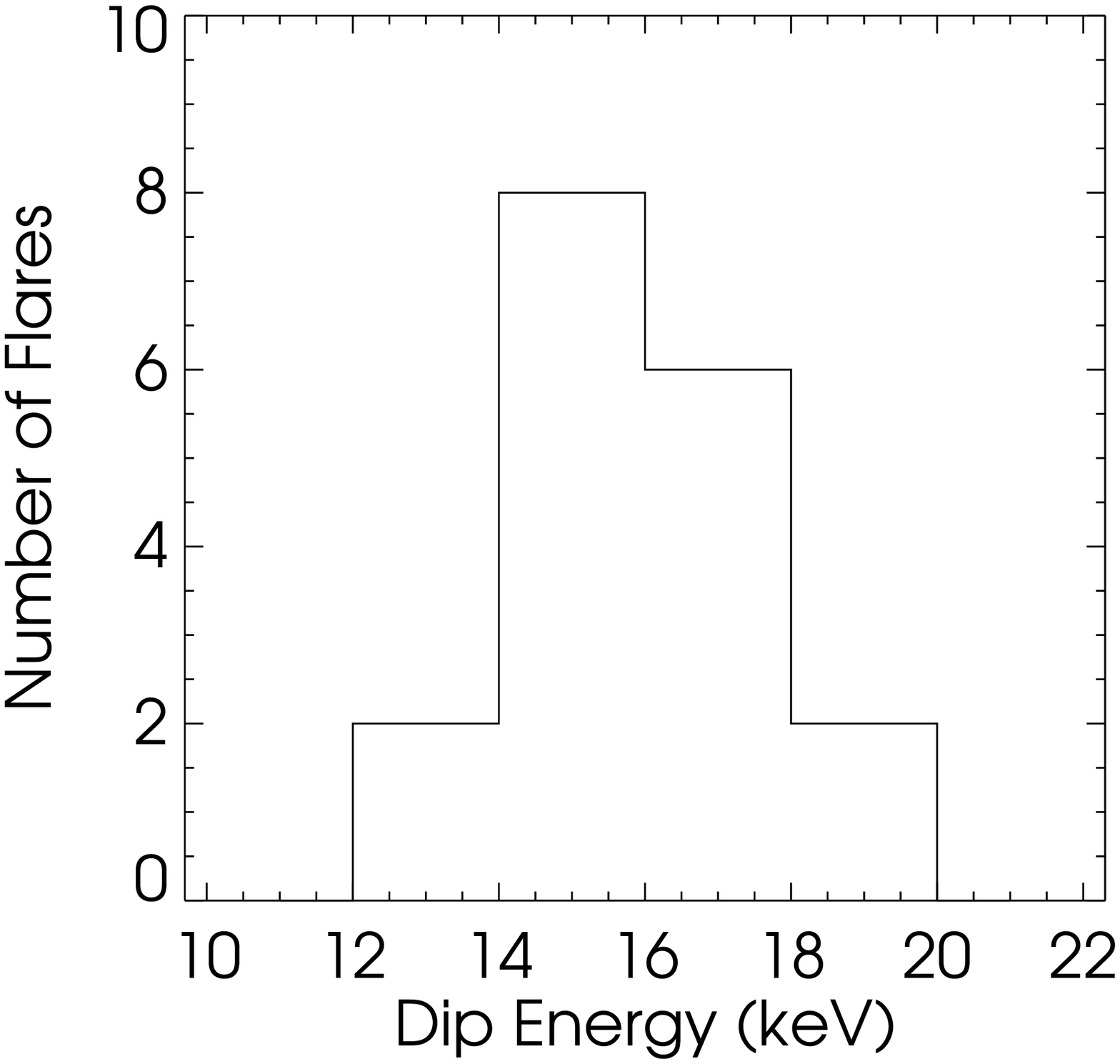}
\end{center}
\caption{Histograms of 18 events with clear dip:
Left panel: Number of events as a function of cosine of heliocentric angle;
Right panel: Number of events as a function of dip
energy $E_{\rm d}$ in keV.}
\label{fig3}
\end{figure}

\begin{figure}
\begin{center}
\includegraphics[width=49mm]{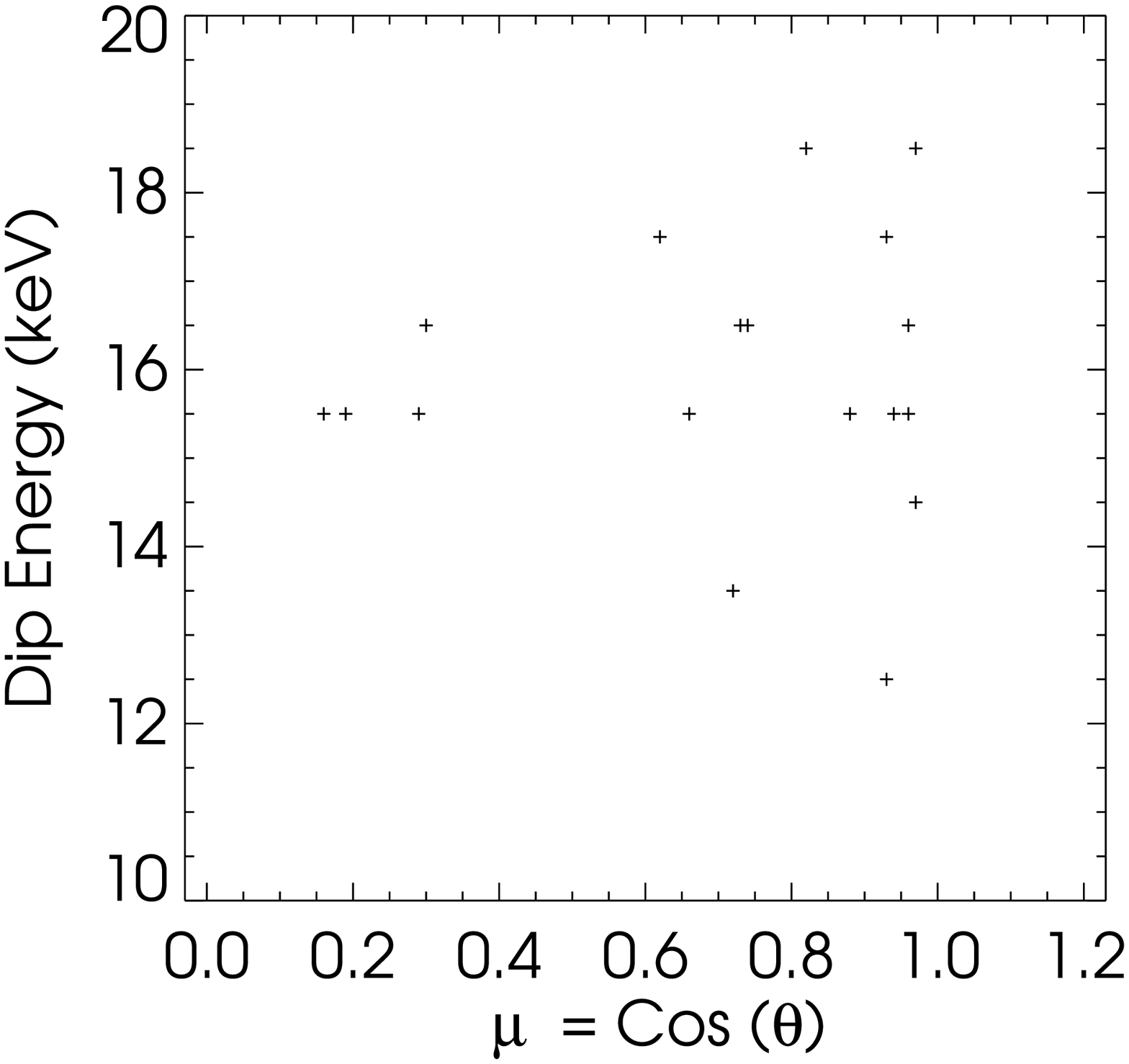}
\includegraphics[width=49mm]{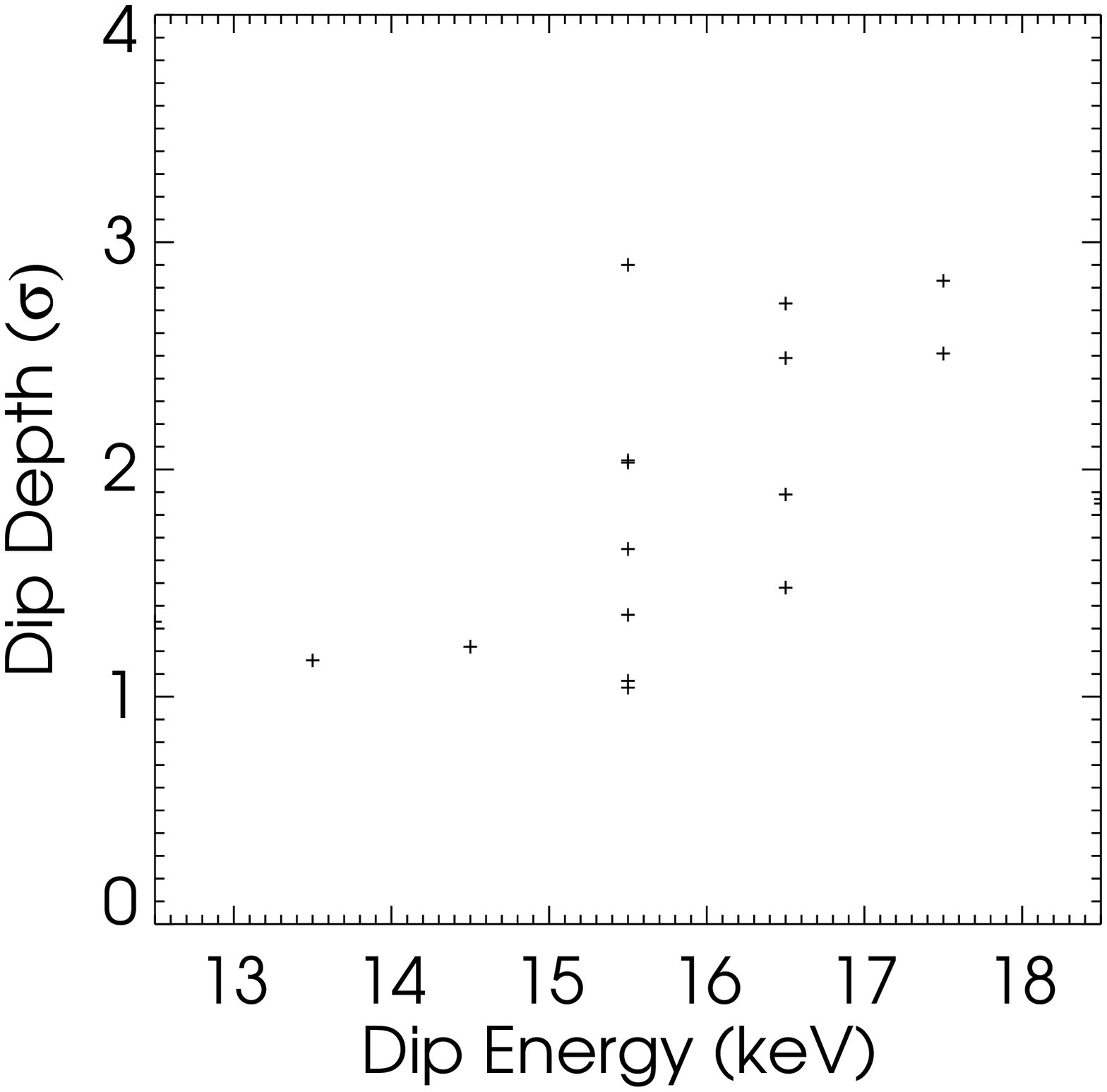}

\end{center}
\caption{Left panel: Dip energy versus cosine of heliocentric angle $\mu$;  Right panel: Dip depth
versus dip energy.}
\label{fig4}
\end{figure}

\begin{figure}
\begin{center}
\includegraphics[width=58mm]{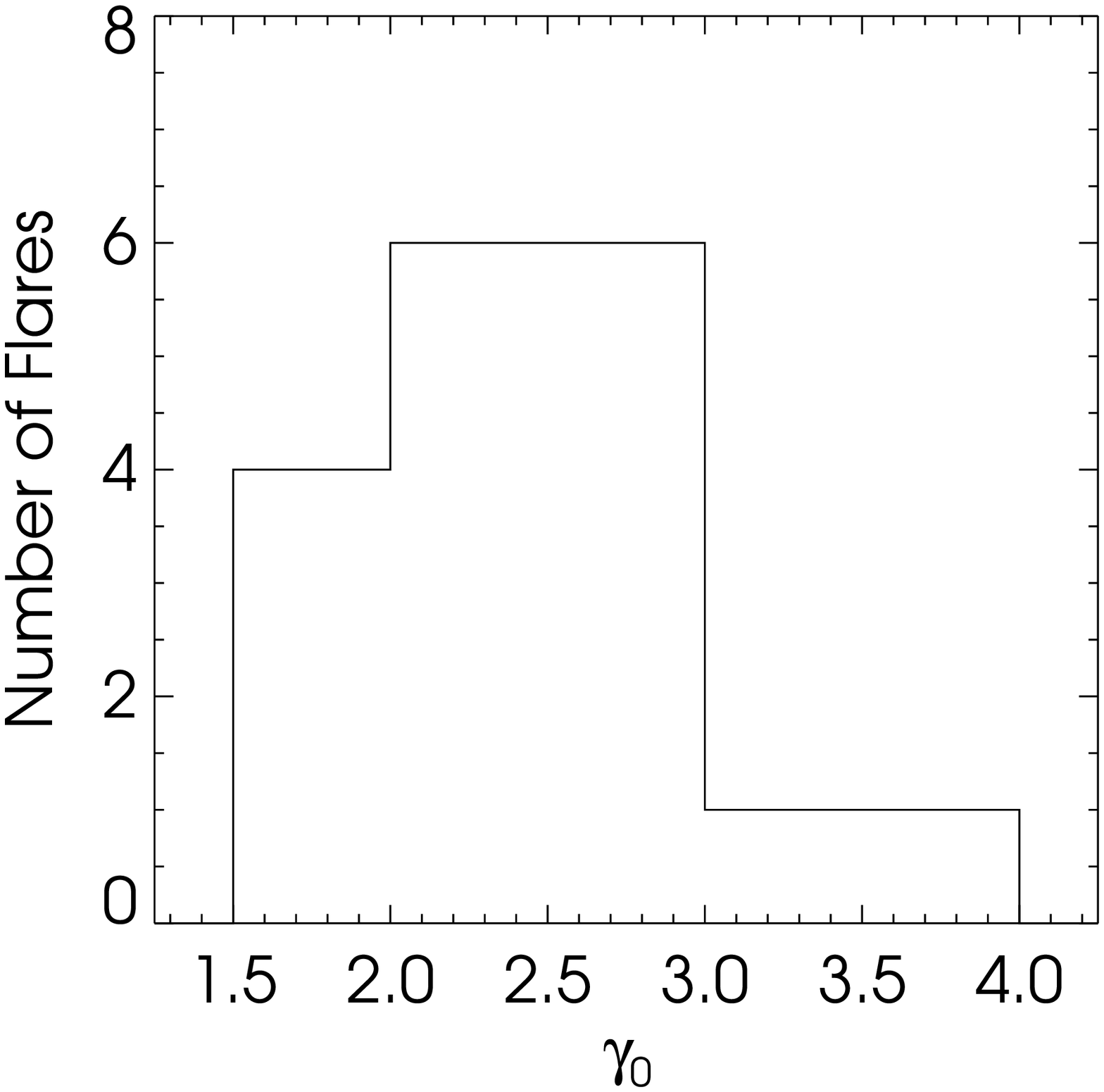}
\includegraphics[width=45mm]{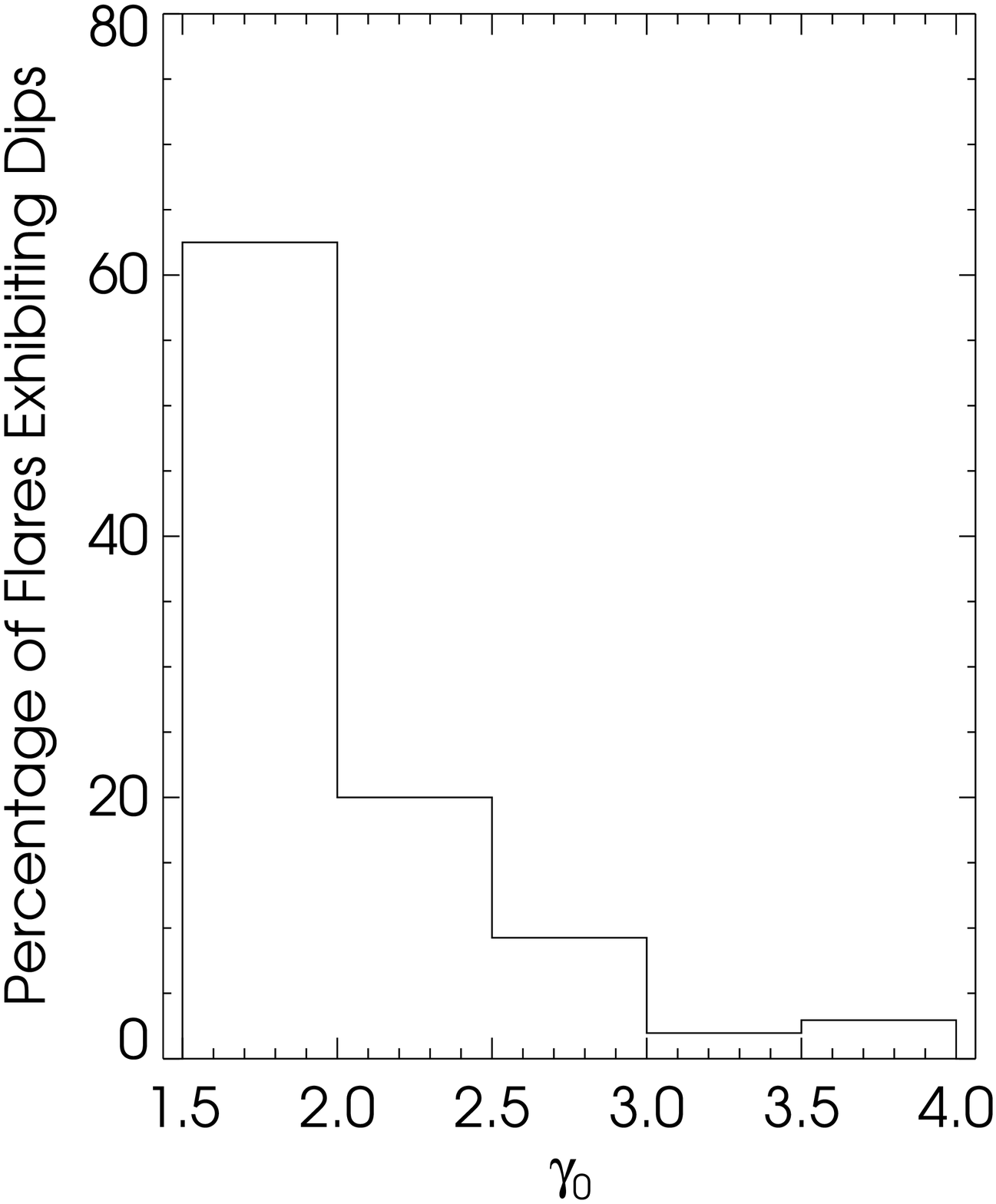}
\end{center}
\caption{Left panel: Histogram of spectral indices $\gamma _0$ for events with a dip;
Right panel percentage of flares exhibiting a dip in the electron spectrum for a given $\gamma_0$.}
\label{fig5}
\end{figure}

The dips are located between the thermal and non-thermal
component and appear approximately at the same energy,
in the range between 13 and 19 keV. The dip energies $E_d$
are given in Table 1 as the bin centre energy. There is no preferential
energy in this range (Figure 3 - right panel.)

There is a clear pattern in the results: flares with dips tend to
occur at locations with large $\mu=\cos\theta$, where $\theta$ denotes
the flare heliocentric angle. Only 4 events are located close to the solar limb
$\mu < 0.5$ while 14 are near the disk centre $\mu < 0.5$ - see left panel
in Figure 3. There is also no strong evidence for the dip energy
being dependent on the flare location or on the dip depth - see left and
right panel in Figure~4, respectively.

\subsection{Correction for X-ray Compton scattered photons}

Previous works \cite{kontar04,kasparova07} have shown that
a feature such as a dip can be a signature of distortion by albedo contribution.
Figure 5 (right panel) shows that larger dips appear for flatter X-ray
spectra. Furthermore, events with large depths tend to appear
close to the disc centre, see Figure~4.
This is consistent with the albedo model \cite{kontar06}
which predicts larger albedo contribution for flat spectra and disc
centre events. It is noteworthy that the albedo contribution
to the observed photon spectrum is still noticeable even for the flares
at heliocentric angles larger than $50^o$. The albedo-corrected mean electron
spectra for flares at $\mu <0.4$ show no dips larger than $1 \sigma$.

The isotropic albedo correction \cite{kontar06} was
applied to all the events with a dip in Table 1 and new ${\bar n}V{\bar F}(E)$,
i.e. corresponding to the primary photon spectra, were derived.
Such albedo corrected mean electron spectra does not reveal any
significant dip, i.e. with depth $\ge 1\sigma$.

\section{Summary and Discussion}\label{summary}

Our analysis shows that the clear dips are rare,
only 18 of 177 events demonstrate a clear dip.
The small number of events with a clear dip or
low energy cutoff can be explained by a variety of reasons.
Firstly, it suggests that the number of very
flat primary spectra is rather small and that the
vast majority of flares have primary spectral index larger
than 2.  Indeed, although the total number of events with a dip
is small (left panel in Figure 5) the fraction of events could
be as high as 60\% for small spectral indices (right panel in Figure 5).
This can be viewed as a lower limit on spectral indices of accelerated
electrons in solar flares. In the case of a thick-target model,
the spectral index of accelerated electrons should be larger than 3.
Secondly, the small number of events with a dip or low energy cutoff
suggests that the thermal component substantially influences the spectrum
in the range of above 10 keV for the majority of flares.
This conclusion is partially supported by \inlinecite{kasparova07}, who
have found a large number of events with very soft spectra with spectral
indices $\gamma _0$ which are larger than 5.

However, when dips occur in the mean electron
spectrum, the local minima in the electron flux spectrum
is consistent with albedo model \cite{kontar06}.
In the standard solar flare model, the electrons are believed to
propagate downwards and hence the reflected flux from the
photosphere should be larger. In this work the albedo was assumed
to be isotropic and this can be viewed as a lower limit
on albedo contribution. Therefore the explanation that albedo
might be overestimated seems unlikely.
As can be seen in Figure 5 flares with a low value of $\gamma_0$
are very likely to exhibit a local minimum in the mean electron
flux spectrum, therefore the small number of flares
with flat spectra results in the the low number of flares with dips.
In addition, the energies of the dip minima are concentrated
near 15 keV, the energy which is expected from
isotropic albedo model (see Figure 1 in \inlinecite{kasparova07}).
We also note that earlier observation of flat X-ray
spectrum are consistent with the albedo model. The
flares suggesting high value of low energy
observed by \inlinecite{nitta90}, \inlinecite{Farnik95}
had flat X-ray spectra and were disk centre events
confirming conclusions of this work.

The low-energy cutoff is often introduced to limit
the total number of non-thermal electrons in solar flares.
Since all dips found in the electron spectra can be easily
"removed" by applying albedo correction, our results allow to conclude
that if low-energy cutoff exists in solar flare spectra it should
be below $\sim 12$ keV. This value puts an upper limit on the
low energy cutoffs and is somewhat less than the values published
in the literature. In addition, since the total number of electrons accelerated
in solar flares is dependent on the low-energy cutoff the lower
value of low-energy cutoff makes the electron
number problem even more severe.

%
\begin{acks}
This work was supported by a PPARC/STFC rolling grant
and Advanced Fellowship (EPK). ED was supported by Cormack summer research scholarship.
JK acknowledges the grant 205/06/P135 of the Grant Agency of the Czech Republic
and the research plan AVZ010030501.
Financial support by the European Commission through the SOLAIRE Network
(MTRN-CT-2006-035484) is gratefully acknowledged by EPK. The authors are thankful
to Sam Krucker for valuable referee comments.
\end{acks}

\end{article}
\end{document}